\documentclass[%
 reprint,
superscriptaddress,
 amsmath,amssymb,
 aps,
 pra,
floatfix
]{revtex4-1}

\usepackage{graphicx}
\usepackage{dcolumn}
\usepackage{bm}
\usepackage{xcolor}
\usepackage{caption}
\usepackage{subcaption}
\usepackage{siunitx}
\usepackage{hyperref}
\hypersetup{hypertexnames=false}
\usepackage{physics}

\setcitestyle{square}

\newcommand{\unit}[1]{\hat{\mathbf{#1}}} 
\renewcommand{\vec}[1]{\mathbf{{#1}}} 

\definecolor{darkblue}{HTML}{000077}
\hypersetup{
	colorlinks = true,
	urlcolor={blue},
	linkcolor={blue},
	citecolor={blue}
}

\usepackage[normalem]{ulem}
\usepackage{color}

\newcommand{\Xout}{\bgroup\markoverwith{\textcolor{red}{\rule[0.5ex]{2pt}{2pt}}}\ULon}

\usepackage{natbib}

\begin{document}

\preprint{APS/123-QED}

\title{The microscopic Einstein-de Haas effect}

\author{T. Wells}
\email{tomos.wells11@imperial.ac.uk}

\author{A. P. Horsfield}%
\email{a.horsfield@imperial.ac.uk}
\affiliation{Department of Materials and Thomas Young Centre, Imperial College London, South Kensington Campus, London SW7 2AZ, United Kingdom}%

\author{W. M. C. Foulkes}
\email{wmc.foulkes@imperial.ac.uk}
\affiliation{Department of Physics and Thomas Young Centre, Imperial College London, South Kensington Campus, London SW7 2AZ, United Kingdom}%
\affiliation{Department of Physics, University of Illinois at Urbana-Champaign, Urbana, IL 61801, USA}%

\author{S. L. Dudarev}
\email{sergei.dudarev@ukaea.uk}
\affiliation{UK Atomic Energy Authority, Culham Center for Fusion Energy, Oxfordshire OX14 3DB, United Kingdom}
\affiliation{Department of Physics and Thomas Young Centre, Imperial College London, South Kensington Campus, London SW7 2AZ, United Kingdom}%

\date{\today}

\begin{abstract}
The Einstein-de Haas (EdH) effect, where the spin angular momentum of electrons is transferred to the mechanical angular momentum of atoms, was established experimentally in 1915. While a semi-classical explanation of the effect exists, modern electronic structure methods have not yet been applied to modelling the phenomenon.  In this paper we investigate its microscopic origins by means of a non-collinear tight-binding model of an O\textsubscript{2} dimer, which includes the effects of spin-orbit coupling, coupling to an external magnetic field, and vector Stoner exchange. By varying an external magnetic field in the presence of spin-orbit coupling, a torque can be generated on the dimer, validating the presence of the EdH effect. Avoided energy level crossings and the rate of change of magnetic field determine the evolution of the spin. We find also that the torque exerted on the nuclei by the electrons in a time-varying $B$ field is not only due to the EdH effect. Other contributions arise from field-induced changes in the electronic orbital angular momentum and from the direct action of the Faraday electric field associated with the time-varying magnetic field.
\end{abstract}

                             
\maketitle

\section{\label{sec:introduction}Introduction}


Materials in burning plasma tokamak fusion devices are expected to be exposed to irradiation in the presence of high strength magnetic fields approaching $\sim $10 T. In addition to stresses that such fields generate in the reactor components \cite{Gade2015}, they may also affect the dislocation microstructure and plasticity of materials \cite{Alshits2003} as well as diffusion of radiation defects. Current tokamak designs, including ITER, use austenitic Fe-Cr-Ni steels \cite{0029-5515-41-3-302} that are non-magnetic on the macroscopic scale while being microscopically antiferromagnetic \cite{Wrobel2015}. In a demonstration fusion reactor (DEMO)~\cite{KOHYAMA19981}, blanket modules are expected to be manufactured from ferromagnetic ferritic-martensitic steels, exhibiting superior resistance to radiation damage. The primary component of a ferritic steel, iron, is strongly ferromagnetic in the body-centered cubic (bcc) phase. The observed directional anisotropy of magnetism in iron \cite{Van_Vleck1937,Daniel2008,Daniel2018} suggests that atomic magnetic moments are coupled to the lattice, giving rise to a magnetic contribution to electric and thermal resistivity \cite{PepperhoffAcet} and to spin-electron-lattice relaxation effects in collision cascades \cite{Ma2012}. 

To help develop the theory of the interaction between electrons and a crystal lattice in a magnetic material in a systematic way, here we consider the Einstein-de Haas (EdH) effect, in which the flipping of spins generates a macroscopic torque on the crystal lattice. This was first predicted in 1908 by O.\ W.\ Richardson and has been demonstrated numerous times since then~\cite{Richardson1908,Barnett1915,Kittel1949}. In Einstein and de Haas's~\cite{EdH} famous experiment, an iron cylinder was suspended by a thin wire inside a solenoid whose axis coincided with that of the cylinder. Varying the magnetic field within the solenoid led to the generation of a measurable torque on the cylinder.

The present-day interpretation of this result is that a change in the external magnetic field changes the magnetization of a ferromagnetic material by realigning the electronic spins. The dimensionless gyromagnetic factor, $g'$, relates the change in  magnetization $\Delta \vec{M}$ to a change in electronic angular momentum $\Delta \vec{J}$ according to
\begin{equation}
	\Delta \vec{M} = -  g' \frac{e}{2m_e} \Delta \vec{J},
\end{equation}
where $e$ is the elementary positive charge and $m_e$ is the mass of an electron.
Conservation of angular momentum requires an equal and opposite change in the mechanical rotational angular momentum of the body of the material. Many experiments have used the EdH effect to measure $g'$ for ferromagnetic materials~\cite{Scott1962}.

The EdH effect relies on the transfer of angular momentum from the electronic spins to the lattice of nuclei and thus highlights the role of spin-orbit coupling (SOC) in spin-lattice interactions~\cite{PhysRevLett.85.3025}. Without SOC, the direction of the electronic spin is decoupled from the orientation of the lattice and the EdH effect does not occur. We note that a good description of SOC is also necessary for the calculation of heat transport coefficients in ferromagnetic materials~\cite{Ohta1982,Ma2012}.

While it may not yet be clear how large an effect magnetism has on thermal conductivity of iron and steels, it should be recalled that the effect on mechanical properties is now known to be significant. For example, nonmagnetic density functional theory (DFT) calculations predict that He interstitials in Fe have the octahedral site as their most favourable position, while magnetic DFT calculations predict that the tetrahedral position is lower in energy~\cite{PhysRevLett.94.046403}. Self-interstitial defects provide another good example: a self-interstitial atom in magnetic bcc iron adopts a $\langle 110 \rangle$ dumbbell structure \cite{Fu2004}, which is different from the linear $\langle 111 \rangle$ crowdion configuration adopted by self-interstitials in non-magnetic bcc transition metals~\cite{PhysRevB.76.054107,PhysRevB.73.020101}. Perhaps even more significantly, magnetic fluctuations give rise to a sequence of $\alpha$-$\gamma$-$\delta$ bcc-fcc-bcc phase transitions in iron and steels at elevated temperature \cite{Lavrentiev2010,Ma2017} that require computing free energy differences between competing magnetic phases with meV accuracy. 

The limitations of our current understanding of magnetic materials are evident from the difficulties in predicting ferromagnetic phenomena such as the sudden collapse of the magnetic moment in hcp-Fe under pressure~\cite{Iota2007}. Lack of understanding also hinders attempts to construct microscopic models of heat flow in ferromagnetic materials~\cite{PepperhoffAcet,Dudarev2013} suitable for use in large-scale atomistic or semi-classical simulations. Our work represents a first step towards addressing this issue.

Although the EdH effect is supported by multiple experiments and illustrates an important way in which magnetic degrees of freedom interact with the lattice degrees of freedom in a material, almost no attempt has been made so far to describe it at a fully quantum-mechanical microscopic level. The authors are unaware of any previous work performing atomic-scale simulations of the EdH effect. However, a few existing studies discuss versions of the EdH effect in a quantum mechanical context. For example, in a numerical model of a pair of dysprosium atoms placed in a spherical harmonic trap interacting through a dipole-dipole interaction term, large orbital angular momenta ($l>20$) were generated by slow variation of an external magnetic field~\cite{0295-5075-116-2-26004}. A recent study of a single-molecule magnet coupled to a nanomechanical resonator constitutes an experimental realisation of the EdH effect for a single molecule~\cite{Ganzhorn2016}.

Magnetic DFT calculations can be expensive to perform and have only rarely been used to study the coupled dynamics of atoms and spins. By contrast, magnetic tight binding (TB) provides a simpler and quicker approach capable of describing much of the same physics. The anomalous iron-chromium mixing enthalpy was explained simply using a fixed-moment model based on TB~\cite{Paxton2008}. Magnetic TB was also used to show that, when magnetic defects are present, the energy surfaces of spin-lattice systems can be highly complex, including multiple separate regions with different spin structures~\cite{Soin2011}.


In this paper we study the EdH effect for an O\textsubscript{2} dimer subject to a time-varying magnetic field. We use a non-collinear TB model that includes coupling to the external magnetic $B$ field, SOC, and vector Stoner exchange~\cite{Coury2016}. Forces on the atoms are calculated using the Hellman-Feynman theorem~\cite{Hellmann1937,PhysRev.56.340}. Once the existence of the EdH effect has been established, we investigate the effect of varying the SOC strength. If SOC is neglected, the component of the spin operator parallel to the applied magnetic field $\vec{B}$ commutes with the Hamiltonian and $\vec{S}\cdot\unit{B}$ is a constant of the motion. For small values of the SOC strength, the EdH torque is correspondingly small, as expected. We find, however, that the EdH effect is not the only torque-generating mechanism experienced by the nuclei due to the electrons under a time-varying $B$ field: the field also couples to the electron orbits, and directly to the nuclei via the Faraday effect. We note that the magnetic TB model used here is extensible to larger systems, including solids.

This paper is structured as follows. Section~\ref{sec:theory} describes the method used for the calculations. The results of the simulations are discussed in section~\ref{sec:results} and conclusions are drawn in section~\ref{sec:conclusions}.
\section{\label{sec:theory}Theory}

\begin{figure}[htp]
	\includegraphics[width=1.0\linewidth]{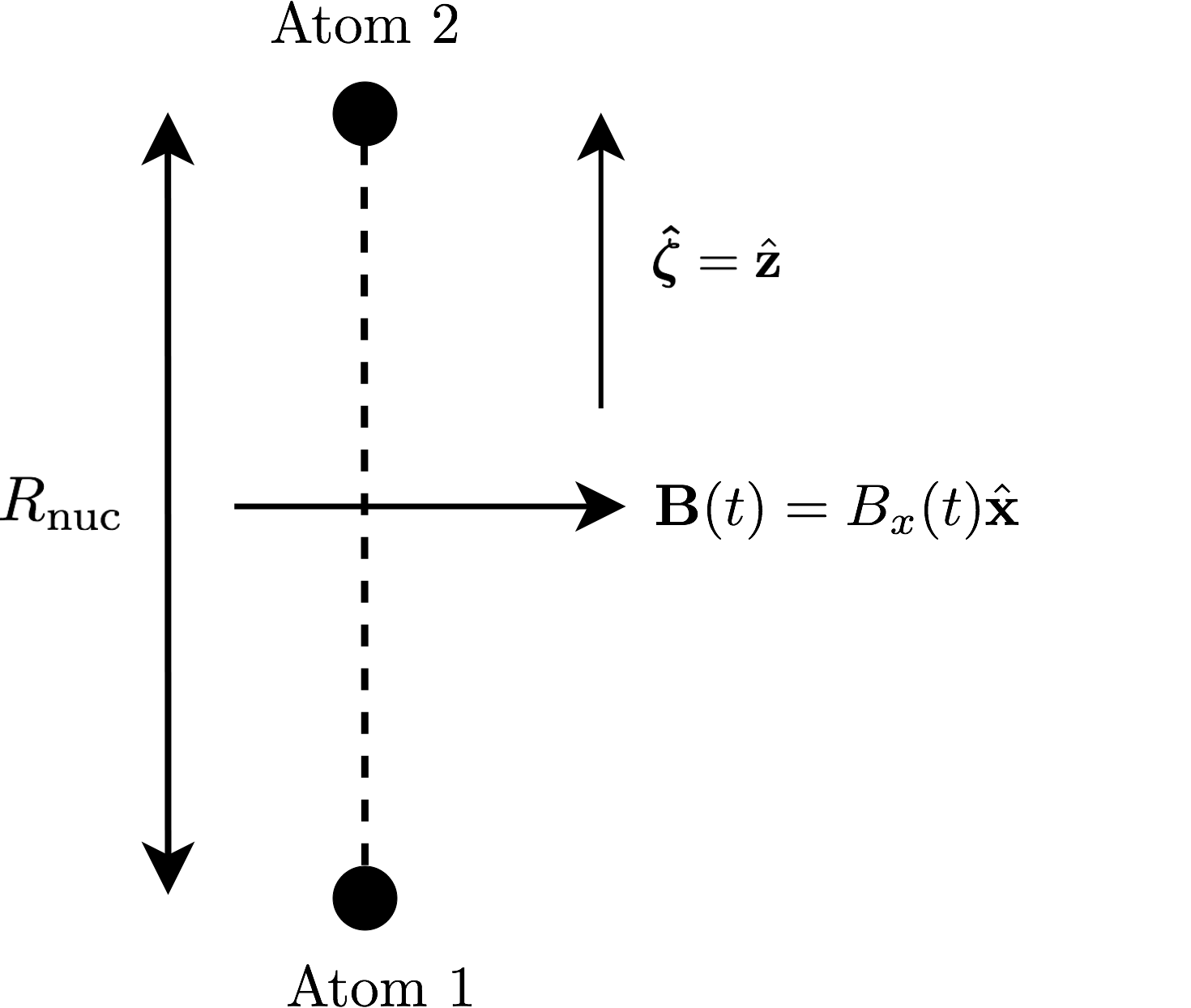}
	\centering
	\caption{Schematic diagram of the oxygen dimer with a time-varying magnetic field applied perpendicular to the dimer axis. The dimer axis is denoted by $\bm{\zeta}$ and $R_\textrm{nuc}$ is the internuclear separation.}
	\label{fig:schematic}
\end{figure}

Fig.~\ref{fig:schematic} shows a schematic diagram of the O\textsubscript{2} dimer studied in this paper. The dimer axis $\hat{\bm{\zeta}}$ is held fixed along the $\unit{z}$ direction. The internuclear separation is $R_{\textrm{nuc}}$. The magnetic field is aligned along $\unit{x}$ and is spatially homogeneous, although its strength may vary in time.
The interaction torque exerted on the nuclei by the electrons is given by
\begin{equation}
  \bm{\Gamma}_\textrm{int} = R_{\textrm{nuc}}(F_{1,y}\unit{x} - F_{1,x}\unit{y} ) = -R_{\textrm{nuc}} (F_{2,y}\unit{x}-F_{2,x}\unit{y}),\label{eq:G_int}
\end{equation}
where atom 1 is below the $xy$-plane, atom 2 is above it, $\bm{F}_{1}$ is the force exerted by the electrons on nucleus 1, and $\bm{F}_{2}$ is the force exerted by the electrons on nucleus 2.

\begin{figure}[htp]
	\centering
	\includegraphics[width=1.0\linewidth]{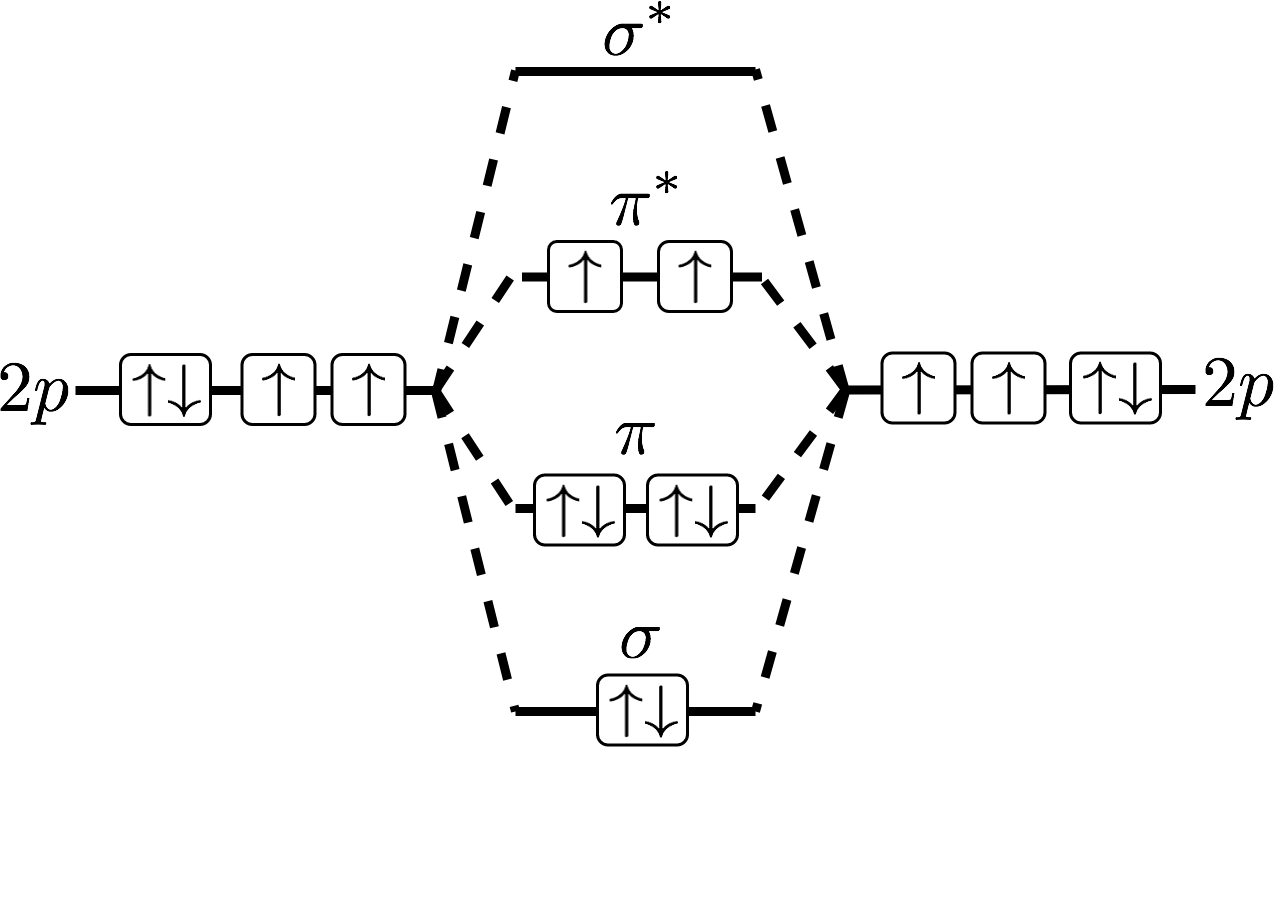}
	\caption{Energy level diagram of the bonding and anti-bonding valence MOs in O\textsubscript{2}. The 1s and 2s shells are fully occupied and are not shown. All of the molecular orbitals appearing in the diagram are derived primarily from atomic $p$ states. The effects of spin-orbit and exchange interactions have been neglected.}
	\label{fig:ox}
\end{figure}

To create a TB model of O\textsubscript{2} it is helpful to know some of its electronic properties. Of the 16 electrons in an O\textsubscript{2} molecule, 4 fill the 1s states, 4 fill the 2s states, and 8 occupy bonding and anti-bonding molecular orbitals (MOs) made from the $2p$ atomic orbitals (AOs). A diagram  of the electronic structure of the $2p$ levels of the molecule is shown in Fig.~\ref{fig:ox}. 

The TB basis functions used here are atomic (or more precisely, atomic-like) $p_x$, $p_y$ and $p_z$ orbitals, with separate orbitals for up and down spins. The six basis functions on each atom are denoted
\begin{equation}
	p_{x,\uparrow}, \; p_{x,\downarrow}, \; p_{y,\uparrow}, \; p_{y,\downarrow}, \; p_{z,\uparrow}, \; p_{z,\downarrow}.
\end{equation}
The basis set does not include the $1s$ and $2s$ orbitals below the $2p$ shell or any orbitals above it, so the dimer Hamiltonian is a $12 \times 12$ Hermitian matrix and the 12 molecular orbitals (MO) are occupied by 8 electrons. The MOs $\phi_n$ are obtained by diagonalizing the Hamiltonian matrix and can be written as linear combinations of the basis functions $\chi_{\alpha\sigma}$ with expansion coefficients $d_{n\alpha\sigma}$:
\begin{equation}
    \phi_n = \sum_{\alpha\sigma} d_{n\alpha\sigma} \chi_{\alpha\sigma},
    \label{eq:expansion}
\end{equation}
where $\alpha$ runs over the spatial AOs on both atoms and $\sigma$ runs over spin states.

Consider first a dimer Hamiltonian containing on-site and hopping matrix elements only, without SOC, exchange interactions, or an applied magnetic field. The
twelve $2p$ MOs are labeled as follows:
\begin{eqnarray}
	\sigma_{\uparrow},~\sigma_{\downarrow},~\pi_{x,\uparrow},~\pi_{x,\downarrow},~\pi_{y,\uparrow}, \pi_{y,\downarrow},\nonumber\\
    \pi^*_{x,\uparrow},~\pi^*_{x,\downarrow},~\pi^*_{y,\uparrow},~\pi^*_{y,\downarrow},~\sigma^*_{\uparrow},~\sigma^*_{\downarrow}, \nonumber
\end{eqnarray}
where the $\sigma$, $\pi_x$ and $\pi_y$ states are bonding combinations of the $p_z$, $p_x$ and $p_y$ AOs, respectively, with $\sigma^*$, $\pi^*_x$ and $\pi^*_y$ the equivalent antibonding states. As shown in Fig.~\ref{fig:ox}, the twelve MOs have only four different energies. The highest occupied shell of molecular spin-orbitals, the $\pi^{*}$ shell, is occupied by 2 electrons.

Although in a purely mean field picture, the 4 $\pi^*$ MOs, $\pi^*_{x,\uparrow}$, $\pi^*_{x,\downarrow}$, $\pi^*_{y,\uparrow}$ and $\pi^*_{y,\downarrow}$, are degenerate, it is known that the ground state of O\textsubscript{2} is a triplet, with the spins of the two $\pi^*$ electrons aligned. The alignment is caused by the exchange interaction, which splits the four degenerate $\pi^*$ spin-orbitals into a three-fold degenerate spin-1 triplet and a spin-0 singlet. Thus a realistic model of ground state O\textsubscript{2} must take the exchange interaction into account.


\subsection{\label{subsec:time_evolution}Time Evolution}

The initial molecular orbitals $\phi_n$ obtained by diagonalizing the Hamiltonian matrix at the beginning of a simulation evolve into time-dependent molecular orbitals $\psi_n(t)$ according to the time-dependent Schr\"{o}dinger equation, $i\hbar\partial_t \psi_n(t) = H(t) \psi_n(t)$, subject to the initial condition $\psi_n(t=0) = \phi_n$. The Hamiltonian $H(t)$ depends on time if the applied $B$ field depends on time, so $\psi_n(t)$ is not in general an exact eigenfunction of $H(t)$ when $t > 0$. The set of time-evolved MOs does, however, remain orthonormal. The time-dependent expansion coefficients $d_{n\alpha}(t)$ are defined by
\begin{equation}
	\psi_n(t) = \sum_{\alpha\sigma} d_{n\alpha\sigma}(t) \chi_{\alpha\sigma}
	\label{eq:time-dependent_expansion}
\end{equation}
and satisfy the following discrete equivalent of the time-dependent Schr\"{o}dinger equation:
\begin{equation}
	i\hbar\frac{\partial}{\partial t} d_{n\alpha\sigma}(t) = \sum_{\alpha'\sigma'} H_{\alpha\sigma, \alpha'\sigma'}(t)d_{n\alpha'\sigma'}(t). \label{eq:eom}
\end{equation}

The method used to evolve this equation in time is time-reversible and unitary. Rewriting the equation of motion, Eq.~\eqref{eq:eom}, with $(D)_{n\alpha\sigma} = d_{n\alpha\sigma}$ and $(H)_{\alpha'\sigma',\alpha\sigma} = H_{\alpha'\sigma',\alpha\sigma}$ as matrices, we obtain
\begin{equation}
	i\hbar\frac{\partial D(t)}{\partial t} = H(t) D(t).
\end{equation}
Letting $dt$ be an infinitesimal positive time interval gives~\cite{Hatano2005}
\begin{equation}
	D(t+dt) = \exp\left({\frac{H(t+\frac{1}{2}dt)}{i\hbar}dt}\right) D(t).
\end{equation}
Thus a finite step of the time evolution can be performed approximately by
\begin{equation}
	d_{n\alpha\sigma}(t+\delta t) =  \sum_{\alpha'\sigma'} \left(e^{H(t+\frac{1}{2}\delta t)\delta t/i\hbar}\right)_{\alpha\sigma,\alpha'\sigma'} d_{n\alpha'\sigma'}(t) .
	\label{eq:rarr}
\end{equation}
The initial values of the expansion coefficients of the time-dependent MO $\psi_n(t)$ are obtained from the corresponding eigenfunction $\phi_n$ of the initial Hamiltonian $H(t=0)$. Since $\psi_n(0) = \phi_n$, we have:
\begin{equation}
   d_{n\alpha\sigma}(0) = \bra{\chi_{\alpha\sigma}}\ket{\phi_n}.
\end{equation}
The time step used in all simulations is 4 Hartree atomic units, which corresponds to approximately \SI{97}{as}.

\subsection{The Hamiltonian}
Describing the EdH effect requires a Hamiltonian that incorporates: (i) the coupling of electrons to a magnetic field; (ii) arbitrary electron spin directions and magnitudes, which may depend on spatial position; and (iii) spin-orbit coupling. To obtain physically meaningful results for O$_2$ it is also necessary to include exchange interactions. A non-collinear TB method can accommodate all of these requirements.

The Hamiltonian employed is
\begin{equation}
  H = H_0 + H_B + H_\textrm{SOC} + H_\textrm{ex},
  \label{eq:ham}
\end{equation}
where $H_0$ is the basic TB dimer Hamiltonian used to obtain Fig.~\ref{fig:ox}, $H_B$ is the coupling of the electrons to the magnetic field, $H_\textrm{SOC}$ is the SOC term, and $H_\textrm{ex}$ is the exchange term.

$H_B$ is taken from the standard Pauli Hamiltonian and is given by $-\bm{\mu}\cdot\bm{B}(t)$, where 
$$\bm{\mu} = -\frac{\mu_B}{\hbar} (\bm{L}+2\bm{S})$$
is the total magnetic moment, $\vec{L}$ and $\vec{S}$ are the canonical orbital and spin angular momentum operators in the Coulomb gauge, $\mu_B$ is the Bohr magneton, and $\bm{B}(t)$ is the external magnetic field acting on the molecule. 

SOC is a relativistic effect and the SOC term in the Hamiltonian may be derived from the Dirac equation via the Foldy-Wouthuysen transformation~\cite{Foldy1950}, which, for spherical potentials, gives
\begin{equation}
	H_{\textrm{SOC}} = \frac{1}{2 m_e^2 c^2}\frac{1}{r} \frac{dV(t)}{dr} \vec{L}\cdot\vec{S},
\end{equation}
where $m_e$ is the mass of an electron, $c$ is the speed of light, and $V(t)$ is the potential experienced by the electron due to the atomic nucleus and the other electrons belonging to that atom in the central field approximation. The gradient of the nuclear potential is largest very close to the nucleus, so the spin-orbit term can be assumed to couple atomic orbitals on the same atom only. The radial part of the SOC matrix element between two orbitals in the same shell on a given atom may then be approximated by a constant given in units of energy, $\xi$, so that \cite{LandauQM}
\begin{equation}
	H_{\textrm{SOC}} \approx \frac{\xi}{\hbar^2} \vec{L}\cdot\vec{S}.
\end{equation}
This is the form of $H_{\textrm{SOC}}$ used in this work.
The next subsection describes the Stoner exchange term, $H_\textrm{ex}$.

\subsection{\label{subsec:stoner}Vector Stoner Exchange}

All of the results reported in this paper have the effects of the vector Stoner exchange term ($H_\textrm{ex}$) included. The more familiar collinear form of the Stoner exchange term is inappropriate for use in describing spin dynamics as it breaks rotational symmetry in spin space~\cite{Coury2016}. The vectorial exchange interaction is treated at the mean-field (Hartree-Fock) level, leading to a self-consistent independent-electron problem.

The remainder of this section outlines the vector Stoner exchange implementation. The exchange moment vector $\vec{m}_a(t)$ on atom $a$ $(\in \{1,2\})$ at time $t$ is evaluated as
\begin{equation}
	\vec{m}_a(t) = \textrm{tr}_{\alpha\in A_a} [\rho(t) \bm{\sigma}] = \textrm{tr} [\rho_a(t) \bm{\sigma}],
\end{equation}
where $\bm{\sigma}$ is the vector of Pauli spin matrices, $\textrm{tr}_{\alpha\in A_a}$ denotes a trace over the atomic orbitals $\alpha$ belonging to atom $a$ only, and $\rho_a(t) \equiv P_a \rho(t) P_a$ is the projection of the time-dependent one-particle density operator
\begin{equation}
    \rho(t) = \sum_{n\;\textrm{occ}} | \psi_n(t) \rangle \langle \psi_n(t) |
\end{equation}
on to the basis of atomic orbitals on atom $a$. We refer to $\vec{m}_a(t)$ as the exchange moment vector in order to differentiate it from the magnetic moment vector $\bm{\mu}$. In terms of expansion coefficients (see Eq.~\eqref{eq:time-dependent_expansion}), the matrix representation of the density operator in the atomic orbital basis is
\begin{equation}
	\rho_{\alpha'\sigma',\alpha\sigma}(t) = \sum_{n\;\textrm{occ}} d^{\phantom{*}}_{n\alpha'\sigma'}(t) d^*_{n\alpha\sigma}(t)
	\label{eq:rho_from_d}
\end{equation}
and the exchange moment of atom $a$ is
\begin{eqnarray}
		\vec{m}_a(t) &=& \sum_{n\;\textrm{occ}} \; \sum_{\alpha\in A_a} \sum_{\sigma^{\prime}, \sigma^{\phantom{\prime}}} d^{\phantom{*}}_{n\alpha\sigma'}(t) d^{*}_{n\alpha\sigma}(t) \bm{\sigma}_{\sigma\sigma'}. \label{eq:m_a}
\end{eqnarray}
From now on, for the sake of notational simplicity, we often suppress the time dependence.

The total energy due to the vector Stoner term is given as
\begin{eqnarray}
	\Delta U = -\frac{1}{4} I \sum_a \vec{m}_a \cdot \vec{m}_a , \label{eq:DeltaU}
\end{eqnarray}
where $I$ is the Stoner exchange parameter given in units of energy.
The TB representation of the self-consistent exchange potential experienced by molecular orbital $n$ is a matrix $\Delta H_{\alpha\sigma,\alpha'\sigma'}$, where
\begin{equation}
	\frac{\partial \Delta U}{\partial d^*_{n\alpha\sigma}} =   \sum_{\alpha'\sigma'} \Delta H_{\alpha\sigma,\alpha'\sigma'}d_{n\alpha'\sigma'}.\label{eq:one}
\end{equation}
For $\alpha\in A_a$ this gives,
\begin{equation}
	\Delta H_{\alpha\sigma,\alpha'\sigma'} = -\delta_{\alpha\alpha'} \frac{1}{2} I \vec{m}_a \cdot \bm{\sigma}_{\sigma \sigma'}. \label{eq:exchange}
\end{equation}

The presence of the Stoner exchange term complicates the time-evolution algorithm described in Sec.~\ref{subsec:time_evolution} (and makes it less than perfectly time reversible) because the Hamiltonian at time $t$, $H(\vec{m}_a(t),t)$, is also a function of the exchange moments at time $t$, which are not calculated in a reversible manner. The evolution of the exchange moments is found by determining the time derivative of the current moment vector $\vec{m}_a(t)$ by a backward difference, $\dot{\vec{m}}_a(t) \approx (\vec{m}_a(t) - \vec{m}_a(t - \delta t))/\delta t$, and Taylor expanding to first order in $\delta t$:
\begin{equation}
      \vec{m}_a(t+\frac{1}{2}\delta t) \approx \vec{m}_a(t) + \frac{1}{2} \dot{\vec{m}}_a(t) \delta t.
\end{equation}
The self-consistency cycle is run only once at the beginning of the calculation when determining the initial set of eigenstates, $\phi_n$. For all subsequent time steps $\vec{m}_a$ is calculated using Eq.~\eqref{eq:m_a}.

\subsection{\label{subsec:model_parameters}Model Parameters}
The TB model requires several experimental parameters. Reference~\cite{Tinkham1955} reports the value of the SOC parameter $\xi$ for an oxygen dimer as $\xi = 2.604~\textrm{meV}$. The same paper gives an experimental value for the bond length $R_\textrm{nuc}$ as \SI{1.21}{\angstrom}. This value is in agreement with ab initio results based on the relativistic Pauli-Breit Hamiltonian~\cite{Fedorov}. The other tight-binding parameters required to describe an O\textsubscript{2} dimer are the Stoner $I$, the on-site energy for $p$ orbitals, $\epsilon_p$, and the hopping parameters, $v_\pi$ and $v_\sigma$. $v_\pi$ is the hopping parameter between $p$ orbitals perpendicular to the dimer axis and $v_\sigma$ the hopping parameter between $p$ orbitals parallel to the dimer axis. The on-site energy used in this work is $\epsilon_p = -16.77$~eV and the hopping parameters are $v_\pi = -0.63 \frac{\hbar^2}{m_e R_{\textrm{nuc}}^2}$ and $v_\sigma = 2.22 \frac{\hbar^2}{m_e R_{\textrm{nuc}}^2}$~\cite{Harrison}. Using the experimental bond length, $R_{\textrm{nuc}} =$ \SI{1.21}{\angstrom}, gives $v_\pi = -3.28$~eV and $v_\sigma = 11.55$~eV. The value of the Stoner $I$ parameter, $I = \SI{0.98}{eV}$, is taken from Ref.~\cite{doi:10.1021/cr010371d}.

\subsection{Calculating Observables}

Forces on the atoms are evaluated using the time-dependent equivalent of the Hellman-Feynman theorem~\cite{Hellmann1937,PhysRev.56.340}. Let $a \in \{1,2\}$ index the two atoms in the dimer and $\vec{R}_a$ denote the position of the nucleus of atom $a$. Using the Hellman-Feynman theorem, we obtain
\begin{align}
	\vec{F}_a &= -\textrm{tr}(\rho \bm{\nabla}_a H ) , \label{eq:force}
\end{align}
for the force on nucleus $a$ due to the electrons, where we have used the definition of the density matrix given in Eq.~\eqref{eq:rho_from_d}. Since our method does not allow the dimer nuclei to move, the nuclei are effectively clamped in position by artificial externally applied forces which oppose any force enacted by the electrons or EM field. The only contributions to the Hamiltonian of Eq.~\eqref{eq:ham} that depend on ${\vec{R}}_1$ and ${\vec{R}}_2$ are the hopping parameters. Using the distance-scaling of the hopping matrix elements discussed in Sec.~\ref{subsec:model_parameters} and Slater-Koster tables~\cite{sutton} for the $p$ orbitals, $\bm{\nabla}_a H$ is calculated analytically.

All the other expectation values computed in this study are found by taking the trace of the operator multiplied by the density matrix, for example, \begin{equation}
	\expval{\vec{L}} = \textrm{tr}(\rho \vec{L}),\ \expval{\vec{S}} = \textrm{tr}(\rho \vec{S}),\ \expval{\bm{\mu}} = \textrm{tr}(\rho \bm{\mu}).
\end{equation}

As shown in Appendix \ref{app:torque}, the classical nuclei experience both an internal torque, $\vec{\Gamma}_{\textrm{int}}$ (defined in Eq.~\eqref{eq:G_int}), exerted by the quantum mechanical electrons, and a direct torque, $\vec{\Gamma}^{N,EM}$, exerted by the applied classical electromagnetic field. The total torque acting on the nuclei is the sum of these two contributions:
\begin{align}
	\vec{\Gamma}^N &= \vec{\Gamma}^{N,EM} + \vec{\Gamma}_\textrm{int}.
	\label{eq:true_mechanical_torque_2}
\end{align}
The electromagnetic torque $\vec{\Gamma}^{N,EM}$ can be expressed as 
\begin{align}
    \vec{\Gamma}^{N, EM} = \sum_{a \in N} \vec{r}_a \cross \vec{F}^{EM}_a ,
 \end{align}
where $N$ is the set of nuclei, $\vec{r}_a$ is the position of nucleus $a$, and
\begin{align}
    \vec{F}^{EM}_a = q_a (\vec{v}_a\times\bm{B}_a) + q_a\bm{E}_a
\end{align}
is the Lorentz force on nucleus $a$. Here $q_a$ is the charge of nucleus $a$, $\vec{B}_a$ and $\vec{E}_a = -\bm{\nabla}_a\varphi_a - \partial_t \vec{A}_a$ are the applied magnetic and electric fields at the position of nucleus $a$, $(\varphi_a/c,\vec{A}_a) = (\varphi(\vec{r}_a,t)/c,\vec{A}(\vec{r}_a,t))$ is the electromagnetic four-potential experienced by nucleus $a$, ${\vec{v}_a =  \frac{1}{m_a}(\vec{p}_a - q_a\vec{A}_a)}$ is the nuclear velocity, $m_a$ is the nuclear mass, and $\vec{p}_a$ is the canonical momentum of nucleus $a$.

We are interested in the dynamics of the nuclei, as these correspond to the motion of the lattice in a solid body. Since the nuclei are held fixed in position in our simulations, they do not experience a force due to the $\vec{v}_a\times\bm{B}_a$ terms in the Lorentz force. The direct EM torque, $\vec{\Gamma}^{N,EM}$, thus arises solely from the circulating electric field produced by the time-varying magnetic field via Faraday's law. The rate of change of the applied magnetic field in our simulations is $10$ T per atomic time unit, implying a direct torque of $-8.9\times 10^{-4} \unit{x} $ Hartree atomic units. This is significant, but our simulations last only 1000 atomic units of time, equivalent to approximately $\SI{2.4e-14}{s}$, while real experiments enact the change in field over times of the order of $\SI{e-2}{s}$~\cite{EdH}. The Faraday torque in our simulations is thus about $10^{12}$ times larger than it would be in a real experiment. Section~\ref{sec:results} reports values of the torque on the nuclei due to the electrons, $\vec{\Gamma}_{\textrm{int}}$. This quantity includes the effects of the Lorentz and Faraday forces on the electrons, which are built in to the Hamiltonian in Eq.~\eqref{eq:ham}, but omits the torque arising from the direct action of the Faraday electric field on the nuclei. If required, the true mechanical torque on the nuclei can be evaluated using Eq.~\eqref{eq:true_mechanical_torque_2}.

\section{\label{sec:results}Results}

To facilitate the interpretation of the physics through a gradual build up in complexity, the discussion of the results is split into two parts. First the results without SOC are presented, followed by the results with SOC. Without SOC we discuss data obtained from simulations with (i) no $B$ field ($\bm{B} = \bm{0}\,$T), (ii) a constant $B$ field of \SI{e3}{T} ($\bm{B} = \num{e3} \hat{\bm{x}}\,$T), and (iii) a linear ramp in $B_x$ from \SI{-5e3}{T} to \SI{5e3}{T} over \SI{e3} Hartree atomic units of time, described by the equation $\bm{B}(t) = (-\num{5e3} + 10t) \hat{\bm{x}}\,$T, where $t$ is measured in atomic units. Such large magnetic fields strengths are used because the magnetic Hamiltonian $H_B = -\bm{\mu}\cdot\bm{B}$ associated with a field of $\SI{1}{T} = \num{4.25e-6}$~atomic units is very small on the atomic energy scale. The decision to reverse the field by applying a linear ramp instead of rotating a $\vec{B}$ vector of constant magnitude was made because the EdH experiment was performed in a magnetic field generated by a solenoid with an oscillatory current. At its center, the solenoid is only capable of producing a magnetic field pointing along its axis. 

The experimental value of the spin-orbit coupling strength in oxygen, ${\xi=2.604}$~meV, is so small that the simulation results with SOC are not visibly different from those without SOC. Thus, in order to observe the EdH effect clearly, it is necessary to use larger values for $\xi$ than are physically realistic for O\textsubscript{2}. In doing so, more can be learned about the effects of SOC on the dimer's dynamics. We consider two different SOC strengths: (i) an intermediate coupling strength of ${\xi=\SI{0.4}{eV}}$; and (ii) a large coupling strength of ${\xi=\SI{e3}{eV}}$. The value of ${\xi=\SI{0.4}{eV}}$ is large enough to have observable effects but not unreasonable for heavier atoms. (Values of $\xi$ as large as $\SI{0.4}{eV}$ have been reported for Sr\textsubscript{2}IrO\textsubscript{4}~\cite{Kim2008}. In bcc Fe the value of $\xi$ is close to 0.06 eV \cite{Desjonqueres2007}.) The unphysically large value of  ${\xi=\SI{e3}{eV}}$ is chosen in order to investigate the limit in which SOC is the dominant energy scale.

Unless stated otherwise, the results below are expressed in Hartree atomic units (a.u.).

\subsection{Without spin-orbit coupling}\label{sec:nosoc}

The results discussed in this section were all obtained in the absence of SOC, i.e., with ${\xi = 0}$~eV. The effects of exchange and the interaction with an external magnetic field are included. The simulations considered have (i) ${B_x=0}\,$T, (ii) ${B_x = \num{e3}}\,$T, and (iii) ${B_x(t)=(-\num{5e3} + 10 t)}\,$T.


%
The $B_x=0\,$T simulation illustrates the behaviour of the electrons in the absence of an external magnetic field.
We find that there is some spin in each of the Cartesian directions. This is because, in the absence of SOC, the spin subsystem is decoupled from the geometry of the dimer which thus does not introduce a preferred direction. Furthermore, since $B=0\,$T, the energy is unaffected by the orientation of the spin: thus the system is completely degenerate with respect to spin direction. The direction of the spin, which is determined by the self-consistency cycle, is therefore the same as the direction of the initial random guess. This arbitrary dependence of the spin on the random direction of the initial guess is a necessary consequence of choosing a spin direction in an otherwise spherically symmetric system. The magnitude of the spin vector is unity, as expected for the triplet ground state of O\textsubscript{2}.

The electrons have zero net orbital angular momentum because the $\sigma$ and $\pi$ subshells are filled, leaving one spin-up electron in each of the two $\pi^*$ orbitals. The total orbital angular momentum is the sum of contributions from the two singly occupied $\pi^*$ orbitals and is thus zero.

The orbital and spin angular momenta $\vec{L}$ and $\vec{S}$ are both constant in time because the Hamiltonian is independent of time and the system is in an energy eigenstate.

In appendix~\ref{app:torque} we show that 
\begin{align}
	\frac{d\expval{\vec{J}}}{dt} = - \bm{\Gamma}_\textrm{int} - \vec{B}\cross\expval{\bm{\mu}},
    \label{eq:dim_torque}
\end{align}
where $\bm{\Gamma}_\textrm{dim}$ is the torque on the dimer and $\vec{J} = \vec{L} + \vec{S}$ is the total canonical electronic angular momentum in the Coulomb gauge. The torque on the dimer is therefore zero in this case since $\vec{B} = \vec{0}$~T and $\vec{L}$ and $\vec{S}$ are constant.


A simulation of the dimer was also performed in the presence of a constant \SI{e3}{T} external magnetic field.
The magnetic field points in the $\unit{x}$ direction and the magnetic field term in the Hamiltonian ($H_B$) is minimized with $\bm{\mu}$ parallel to $\bm{B}$, so the spin lies anti-parallel to the $\vec{B}$ field.

The orbital angular momentum, which is non-zero, also points opposite to the $B$ field and remains constant in time, again because the Hamiltonian is independent of time and the system is in an energy eigenstate.
As is expected for this static system, Eq.~\eqref{eq:dim_torque} shows that the torque is zero.

\begin{figure}[htp]
	\centering
   	\begin{subfigure}{.5\textwidth}
		\centering
		\includegraphics[width=1.0\linewidth]{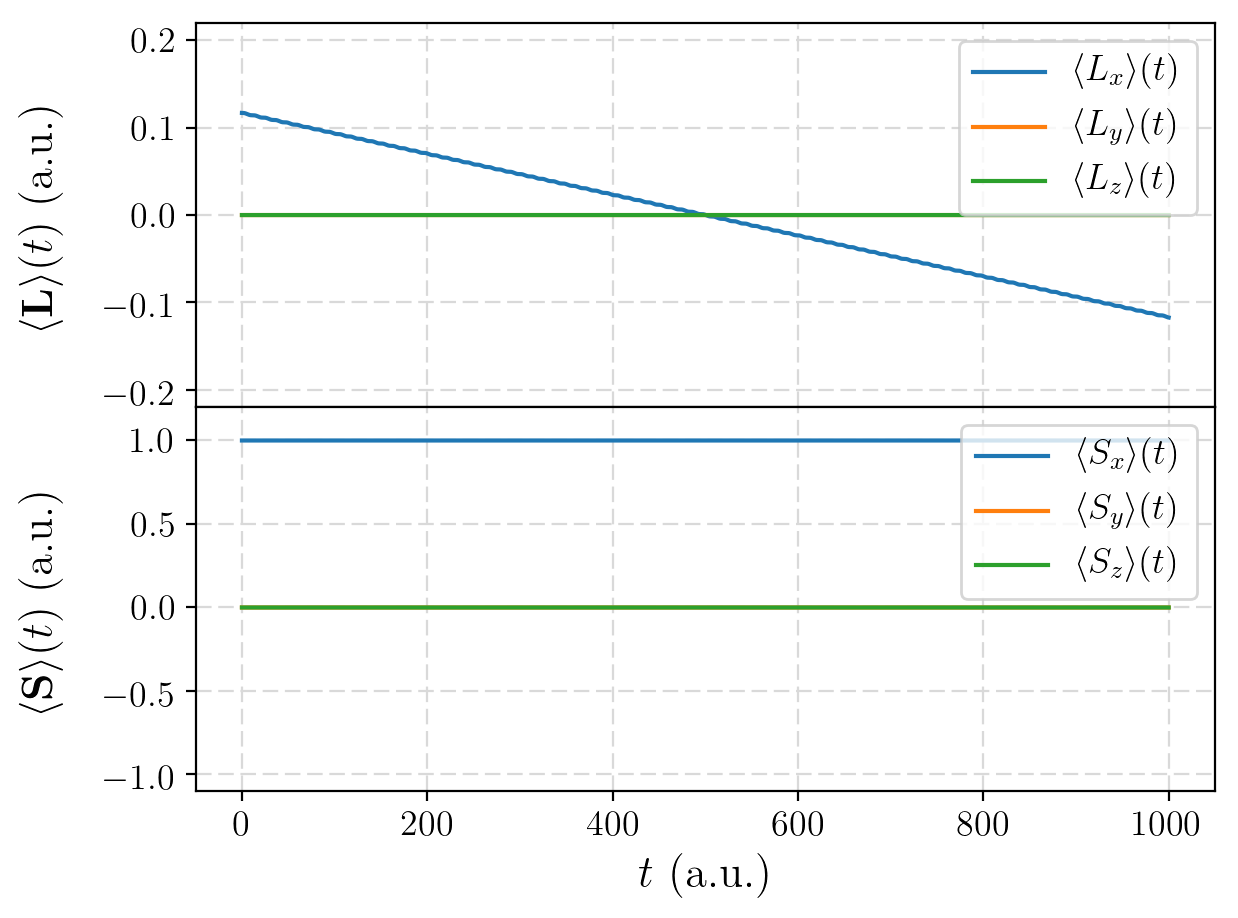}
		\caption{}
		\label{fig:NoSOC_B=-5000+10t_1}
	\end{subfigure}%
    \vskip\baselineskip
	\begin{subfigure}{.5\textwidth}
		\centering
		\includegraphics[width=1.0\linewidth]{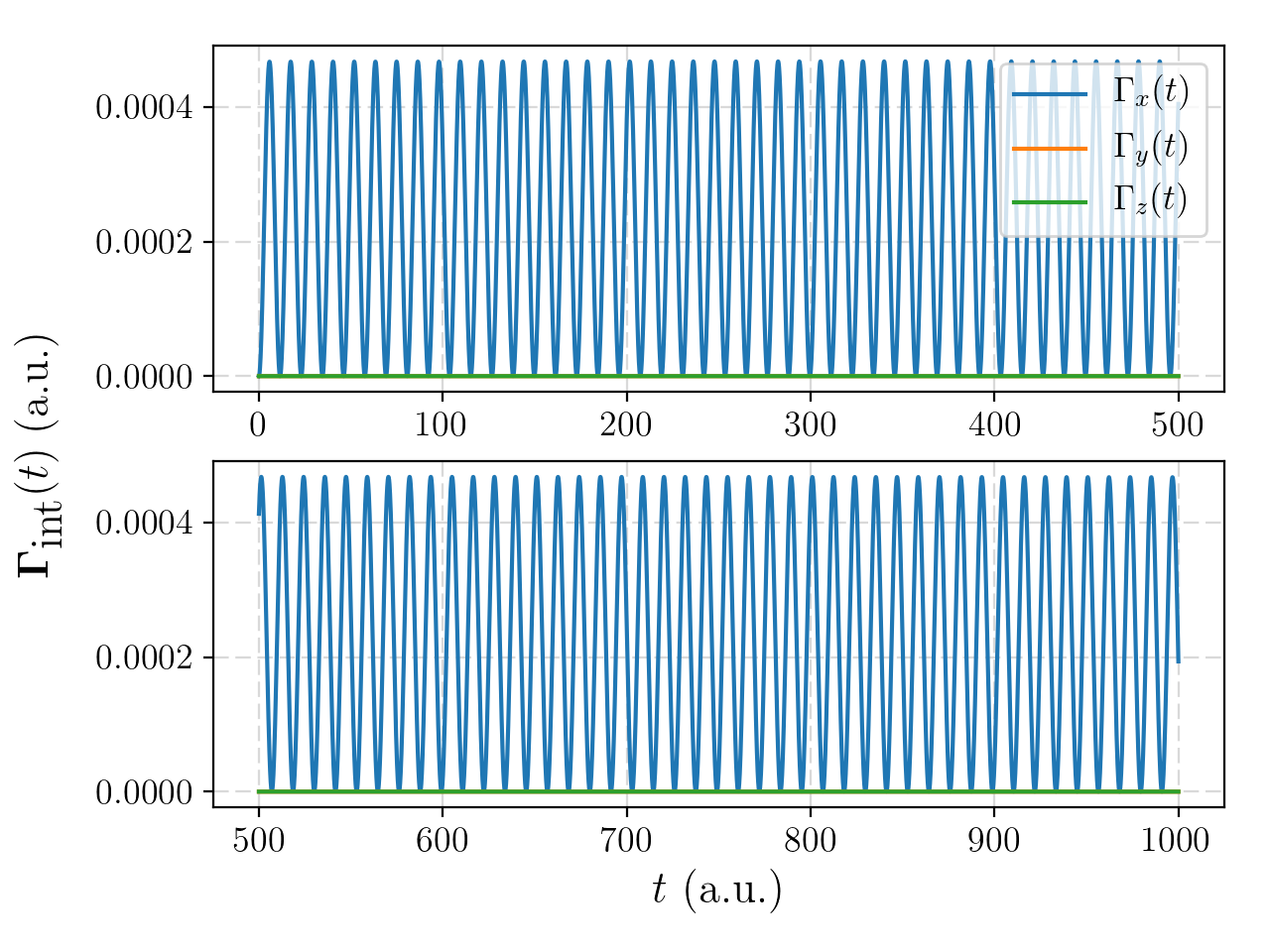}
		\caption{}
		\label{fig:NoSOC_B=-5000+10t_2}
	\end{subfigure}
    \caption{Time-evolution of (a) angular momenta and (b) torque expectation values when the applied magnetic field varies with time: $B_x(t) = \num{-5e3}+10t$~T. The spin is constant and anti-parallel to the initial $B$ field, but the orbital angular momentum remains almost proportional to $-\vec{B}$ throughout the simulation, with some additional oscillations due to the coupling caused by $L_x$. The dimer experiences an oscillatory torque with a non-zero average value. The $y$ components of $\expval{\vec{S}}(t)$, $\expval{\vec{L}}(t)$ and $\vec{\Gamma}(t)$ are zero throughout the simulation and the orange lines representing them are hidden below the green lines representing the $z$ components.} 
	\label{fig:NoSOC_B=-5000+10t}
\end{figure}

Fig.~\ref{fig:NoSOC_B=-5000+10t} shows the simulation results for the magnetic field profile $\vec{B}(t) = B_x(t)\unit{x} = (\num{-5e3}+10t)\unit{x}$. This corresponds to a linear ramp from \SI{-5e3}{T} to \SI{5e3}{T} over a duration of \SI{e3}a.u.
The spin stays constant, pointing in the $\unit{x}$ direction throughout. The applied field $B_x$ is negative at $t=0$ and the spin is oriented antiparallel to this field in the initial ground state. The spin remains constant as it is initially aligned along $\unit{x}$ and $S_x$ commutes with the Hamiltonian. The spin is unable to flip in response to the reversal of the applied magnetic field.

The evolution of the orbital angular momentum over time has two notable features. The first is an adiabatic effect in which the amount of orbital angular momentum pointing along the $B$ field is proportional to minus the field. The second effect is a small sinusoidal variation, barely visible in Fig.~\ref{fig:NoSOC_B=-5000+10t_1}, caused by a Rabi oscillation between eigenstates split by the $B$ field. Section~\ref{sec:osc_freq} explains this effect by deriving the energy difference of the splitting and thus the frequency of the oscillations.
A torque is exerted on the dimer by the change in orbital angular momentum of the electrons. Although the torque is oscillatory, its time average is non-zero. The dimer would therefore start to rotate if its nuclei were not clamped in position. This torque is not related to the EdH effect since $-{d\expval{\vec{S}}}/{dt} = \vec{0}$ and there is no spin-orbit coupling.

\subsection{Oscillations induced by the time-dependent magnetic field}
\label{sec:osc_freq}

An analytic expression for the oscillation frequency of the orbital angular momentum can be found by considering the form of $H_B = -\bm{\mu}\cdot\bm{B}$ in the MO basis. Let $(M^T)_{n\alpha\sigma} \equiv d_{n\alpha\sigma}$ be the transformation matrix from the AO basis to the time-independent MO basis at $B = 0$~T. The expansion coefficients $d_{n\alpha\sigma}$ are as defined in Eq.~\eqref{eq:expansion}. The zero-field MOs $\phi_n$ are given by $\phi_n = (M^T)_{n,\alpha\sigma} \chi_{\alpha\sigma}$, where $\chi_{\alpha\sigma}$ is an AO and repeated suffices are summed. The matrix representations of the angular momentum operators transform as
\begin{align}
	\tilde{\bm{L}}_{mn} &= (M^\dagger \bm{L} M )_{mn}, \\
    \tilde{\bm{S}}_{mn} &= (M^\dagger \bm{S} M )_{mn},
\end{align}
where $\bm{L}$ and $\bm{S}$ are matrices in the atomic orbital basis and $\tilde{\bm{L}}$ and $\tilde{\bm{S}}$ are matrices in the $B=0$ MO basis. The $B=0$ Hamiltonian $H_0$ is diagonal in the MO basis, so only the Zeeman term $-\bm{\mu}\cdot\bm{B} = -\mu_x B_x$ mixes MOs when $B_x \neq 0$. By quantizing the spins along $\unit{x}$ rather than $\unit{z}$, so that $\tilde{S}_x$ is diagonal, we ensure that the only off-diagonal contributions to $-\mu_x B_x \sim (L_x + 2 S_x)B_x$ are those arising from the matrix $\tilde{L}_x$.

In this form the $12 \times 12$ Hamiltonian matrix becomes block diagonal, consisting of four uncoupled states ($\pi_{x\uparrow}$, $\pi_{x\downarrow}$, $\pi^*_{x\uparrow}$ and $\pi^*_{x\downarrow}$), and four $2\times 2$ blocks, each of which couples one of the following pairs of MOs:
\begin{enumerate}
	\item $\sigma_{\uparrow}$ and $\pi^*_{y\uparrow}$,
	\item $\sigma_{\downarrow}$ and $\pi^*_{y\downarrow}$,
	\item $\pi_{y\uparrow}$ and $\sigma^*_{\uparrow}$,
	\item $\pi_{y\downarrow}$ and $\sigma^*_{\downarrow}$.
\end{enumerate}
For example, the $2\times 2$ matrix describing the coupling between $\sigma_{\uparrow}$ and $\pi^*_{y\uparrow}$ is
\begin{equation}
  \begin{pmatrix}
    \epsilon_{\sigma}+B_x/2 & -iB_x/2 \\
    iB_x/2 & \epsilon_{\pi^*}+B_x/2
  \end{pmatrix},
\end{equation}
which has eigenvalues
\begin{equation}
  \epsilon_{1,\pm} = \frac{1}{2}\bigg(\epsilon_\sigma+\epsilon_{\pi^*}+B_x \pm \sqrt{(\epsilon_\sigma-\epsilon_{\pi^*})^2+B_x^2(t)}\bigg).
\end{equation}
The magnitude of the difference between the two eigenvalues is
\begin{equation}
  \Delta\epsilon_1 = \sqrt{(\epsilon_\sigma-\epsilon_{\pi^*})^2+B_x^2(t)}.
\end{equation}
This energy difference corresponds to a Rabi angular frequency $\omega_1 = \Delta \epsilon_1 / \hbar$ and hence to an oscillation in $\expval{L_x}$. The magnitude of $\Delta \epsilon_1$ varies slowly as the applied B field varies.

To check that this explanation accounts for the observed oscillations in $\expval{L_x}$,  consider the case when ${B_x = \num{5e3}}$~T, for which $\Delta\epsilon_1 = 0.54552$~Ha and $T = 1/f = h/\Delta\epsilon_1 = 11.52$~a.u. This agrees well with the oscillation period seen in the simulation results in Fig.~\ref{fig:NoSOC_B=-5000+10t}. The period associated with the other three pairs of coupled MOs is the same.

\subsection{With spin-orbit coupling}\label{sec:soc}

Since the results of the simulations without SOC are practically indistinguishable from results obtained using the small value of $\xi$ appropriate for a real oxygen atom, two further simulations were carried out, one for an intermediate value of $\xi$ and another for a very large value of $\xi$ intended to approach the limit in which SOC is the dominant energy scale.

\begin{figure}[htp]
	\centering
   	\begin{subfigure}{.5\textwidth}
		\centering
		\includegraphics[width=1.0\linewidth]{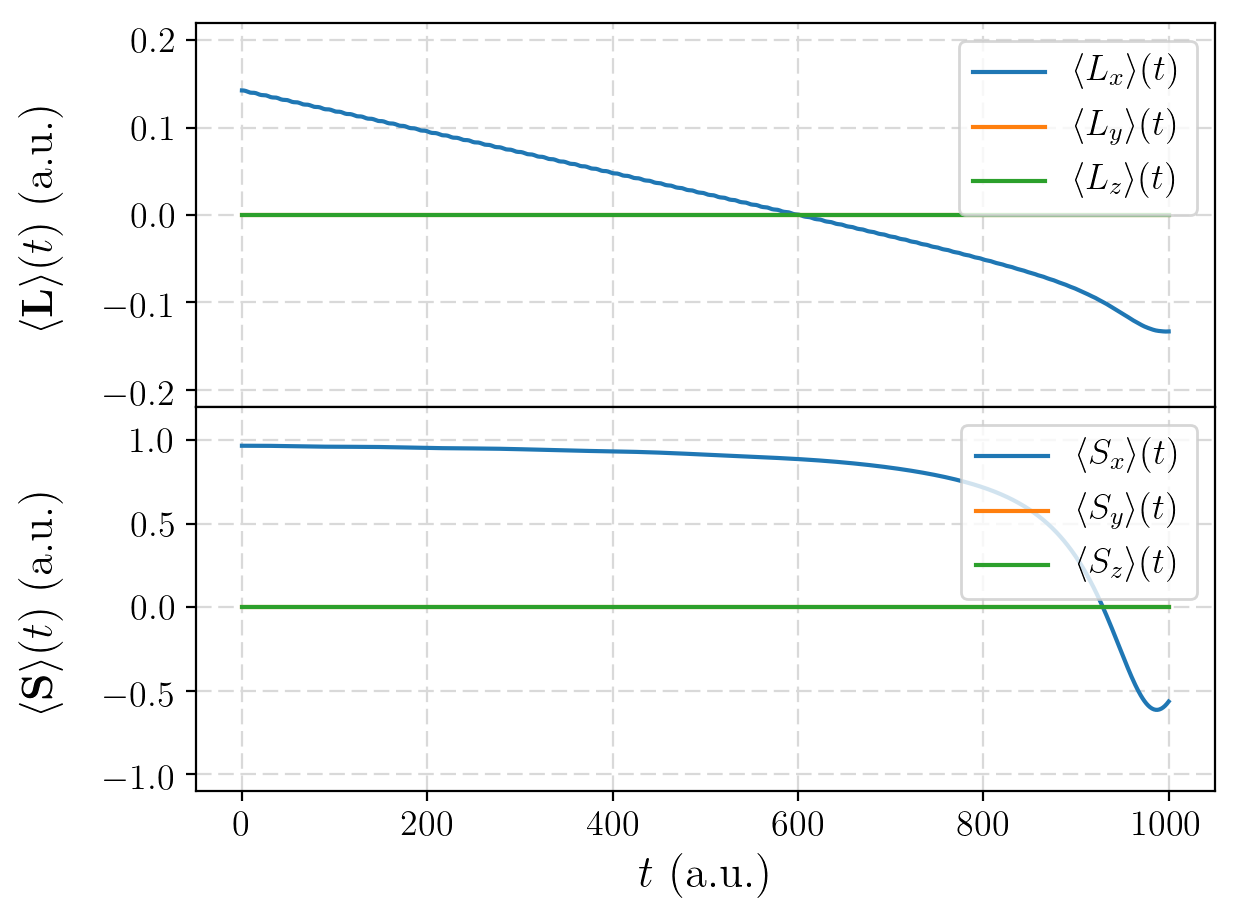}
		\caption{}
		\label{fig:intermediateSOC_B=-5000+10t_1}
	\end{subfigure}%
    \vskip\baselineskip
	\begin{subfigure}{.5\textwidth}
		\centering
		\includegraphics[width=1.0\linewidth]{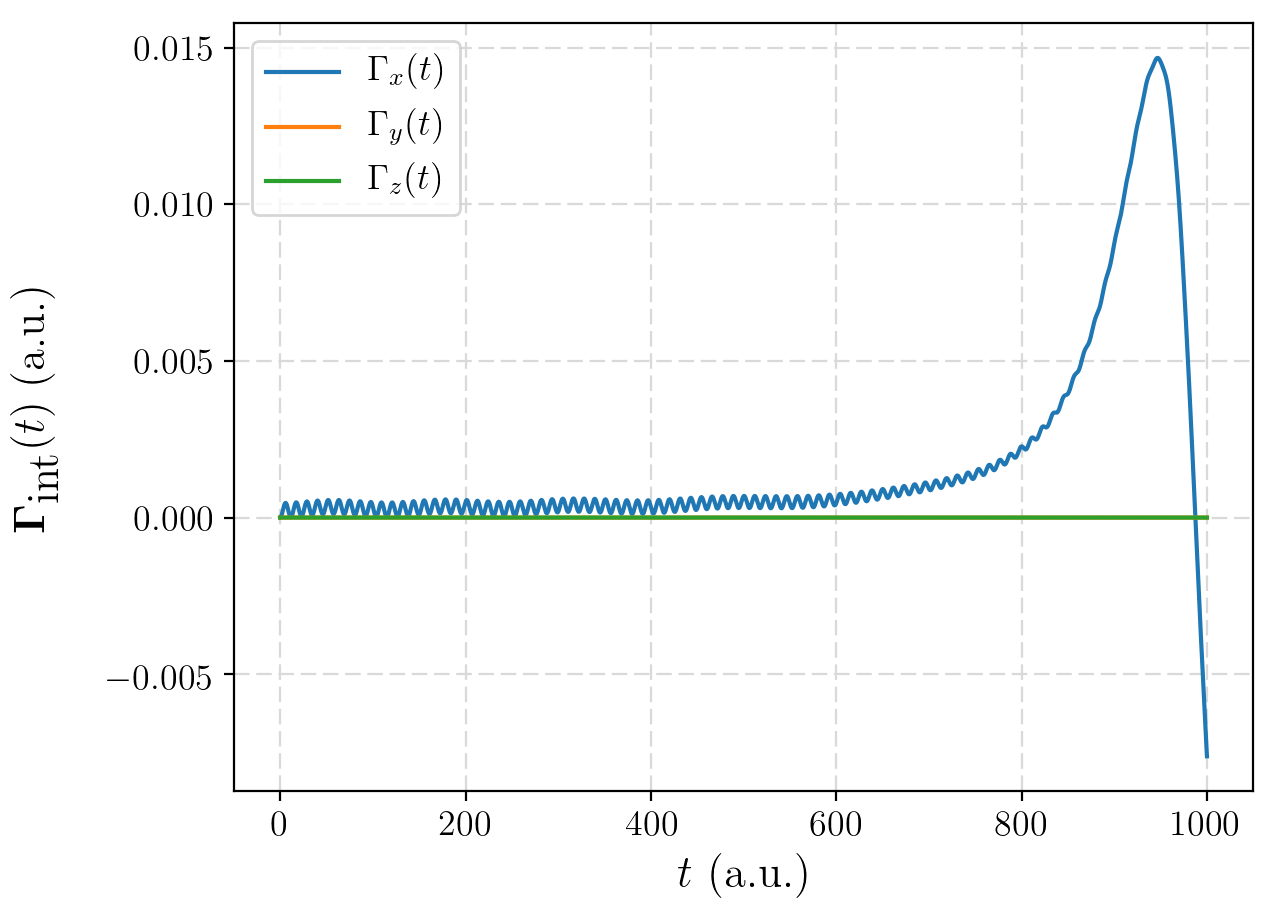}
		\caption{}
		\label{fig:intermediateSOC_B=-5000+10t_2}
	\end{subfigure}
    \caption{Time-evolution of (a) angular momenta and (b) torque expectation values with $B(t) = \num{-5e3}+10t$~T and $\xi=0.4$~eV. The spin evolves gradually at first but towards the end of the simulation it changes sign; this causes a corresponding change of sign in the direction of the torque. The orange lines are hidden beneath the green lines.}
	\label{fig:intermediateSOC_B=-5000+10t}
\end{figure}

%
Fig.~\ref{fig:intermediateSOC_B=-5000+10t} shows results from a simulation with ${\xi=0.4}$~eV and a linear ramp in the $B$ field.
At the beginning of the simulation the dynamics look similar to the SOC-free behaviour shown in Fig.~\ref{fig:NoSOC_B=-5000+10t}. The most noticeable difference is a slight decrease in the magnitude of the initial value of $\expval{L_x}$.
Since $\hat{S}_x$ no longer commutes with the Hamiltonian when SOC is included, its expectation value can now change with time as demonstrated in Fig.~\ref{fig:intermediateSOC_B=-5000+10t_1}. Towards the end of the simulation the spin changes sign, which leads to the change in direction of the torque shown in Fig.~\ref{fig:intermediateSOC_B=-5000+10t_2}. Thus, unlike the SOC-free case, the spin is able to reverse direction in response to the reversal of the applied field. The spin flip takes place near $t = 800$ a.u., well after the time, $t = 500$ a.u., at which the applied field passes through zero.

\begin{figure}[htp!]
	\centering
   	\begin{subfigure}{.5\textwidth}
		\centering	\includegraphics[width=1.0\linewidth]{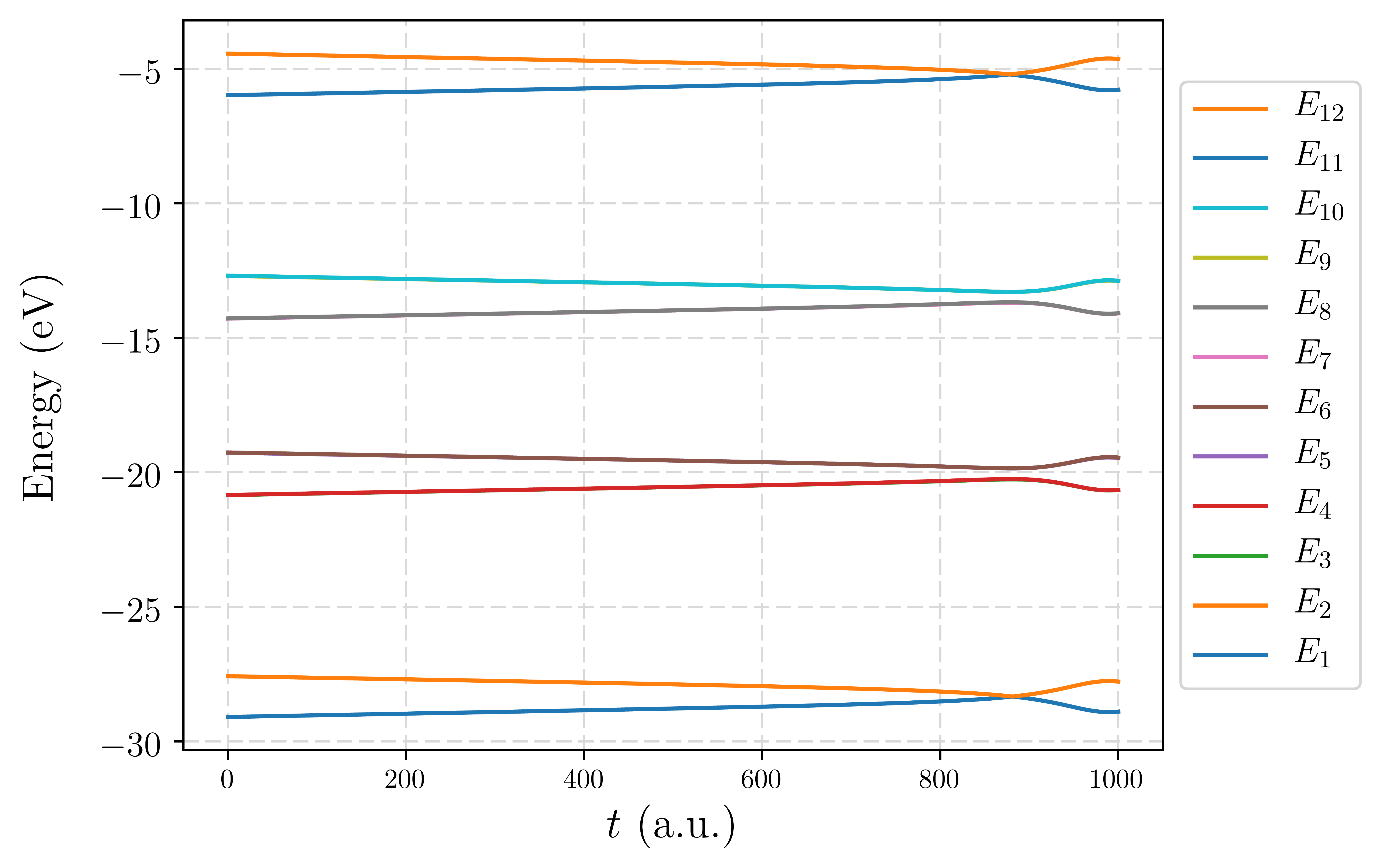}
	\end{subfigure}%
    \caption{Evolution of the energies of the instantaneous molecular orbitals of the time-evolved Hamiltonian. Energy levels 1 and 2 correspond to the $\sigma$ bonding orbitals; energy levels 3--6 (each doubly degenerate) to the $\pi$ bonding orbitals; energy levels 7--10 (again doubly degenerate) to the $\pi^{\ast}$ anti-bonding orbitals; and energy levels 11 and 12 to the $\sigma^{\ast}$ anti-bonding orbitals. The crossing of energy levels (an avoided crossing in the case of the $\pi$ and $\pi^{\ast}$ orbitals) occurs at approximately $t=800$~a.u., when the $-\bm{\mu}\cdot\vec{B}$ term in the Hamiltonian matches and counters the exchange splitting. The crossings of $\pi$ and $\pi^{\ast}$ orbitals are avoided since these states are split by SOC, whereas the $\sigma$ and $\sigma^{\ast}$ states undergo true crossings, as they are not split by SOC.}
	\label{fig:TEMO}
\end{figure}

Figure~\ref{fig:TEMO} shows the eigenvalues of the time-evolved Hamiltonian, $H(\vec{m}_a(t),t)$, as a function of time. These `instantaneous' eigenvalues are calculated by diagonalizing $H(\vec{m}_a(t),t)$ non-self-consistently. Because $H(\vec{m}_a(t),t)$ depends on $\vec{m}_a(t)$, which evolved from previous time steps, the instantaneous eigenvalues and eigenstates are history dependent. Stoner exchange thus introduces a memory into the system. The energy level crossings near $t =800$~a.u.\ determine the timing of the spin flip.

It is important to distinguish the instantaneous eigenvalues shown in Fig.~\ref{fig:TEMO} from the self-consistent eigenvalues at time $t$ (not shown), which are obtained by carrying out a fully self-consistent diagonalization of the Hamiltonian in the presence of a fixed applied magnetic field $B(t)$. Because the magnetic moments are always at their self-consistent values, the self-consistent eigenvalues are not history dependent. Neither the instantaneous eigenfunctions nor the self-consistent eigenfunctions are the same as the time-evolved molecular orbitals.

For $B_x < 0$~T, the lowest energy direction for $\vec{m}_a$ is to be aligned along $\unit{x}$ since this minimizes the spin contribution to $-\bm{\mu}\cdot\vec{B}$. As $B_x(t)$ rises through zero, the new minimum energy direction for $\vec{m}_a$ becomes $-\unit{x}$. Thus, if a simulation were performed in which the system were solved self-consistently at each time step, an energy level crossing would occur at $t=500$~a.u., when the magnetic field $B_x(t)$ passed through zero.

In a time-evolved simulation, $\vec{m}_a(t)$ is calculated from its previous values and the $\frac{1}{2} I \vec{m}_a \cdot \bm{\sigma}_{\sigma\sigma'}$ vector Stoner exchange term in the effective Hamiltonian acts to keep $\vec{m}_a(t)$ aligned along its current value. A lower energy state with all spins reversed exists, but an energy barrier has to be surmounted to reach it. This causes $\vec{m}_a(t)$ to maintain its original direction for longer than in the corresponding self-consistent calculation, explaining the delay in the reversal of the spin of the time-evolved Hamiltonian shown in Fig.~\ref{fig:TEMO}. Additional simulations show that the time at which the avoided crossing occurs moves towards $t=500$~a.u.\ as the Stoner exchange parameter $I$ is reduced and can be made to occur later by increasing $I$.

\begin{figure}[htp!]
	\centering
   	\begin{subfigure}{.5\textwidth}
		\centering	\includegraphics[width=1.0\linewidth]{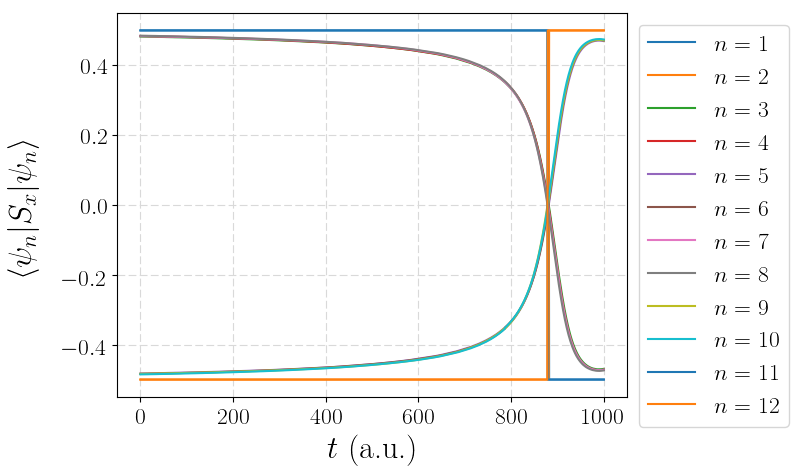}
	\end{subfigure}%
    \caption{Evolutions of the spins of the time-dependent molecular orbitals $\psi_n(t)$ over time. The change in spin direction occurs slightly after the crossing of energy levels shown in Fig.~\ref{fig:TEMO}. The total spin expectation value $\expval{S_x}$ is the sum of the expectation values for the occupied orbitals, $n=1$ to $n=8$.} 
	\label{fig:s_x}
\end{figure}
Comparing Figs.~\ref{fig:TEMO} and \ref{fig:s_x}, we see that the reversal in the sign of the spin of the time-dependent $\pi$ and $\pi^{\ast}$ molecular orbitals does not occur at the same time as the avoided crossing, but slightly afterwards. The adiabatic theorem says that the evolution of a time-dependent MO closely follows the evolution of the corresponding eigenstate of the Hamiltonian if the rate of change of $B$ is small enough. The electron therefore remains predominantly in the adiabatically connected eigenstate of the time-evolved Hamiltonian, which changes its spin direction as the avoided crossing is traversed. Thus, the delay in the rotation of the spins of the time-evolved molecular orbitals as they respond to the change in the time-evolved Hamiltonian is due to the simulation being diabatic, which increases the tendency of the time-evolved MO to remain similar to its original instantaneous MO of the time-evolved Hamiltonian during the avoided crossing. This is supported by further simulations in which $\dot{B}$ is varied and it is found that for low $\dot{B}$, the change in sign of the spin of the time-dependent $\pi$ and $\pi^{\ast}$ molecular orbitals tends toward being simultaneous with the change in sign of the spin of the $\pi$ and $\pi^{\ast}$ instantaneous eigenstates of the time-evolved Hamiltonian. 

\begin{figure}[htp!]
	\centering
   	\begin{subfigure}{.5\textwidth}
		\centering
		\includegraphics[width=1.0\linewidth]{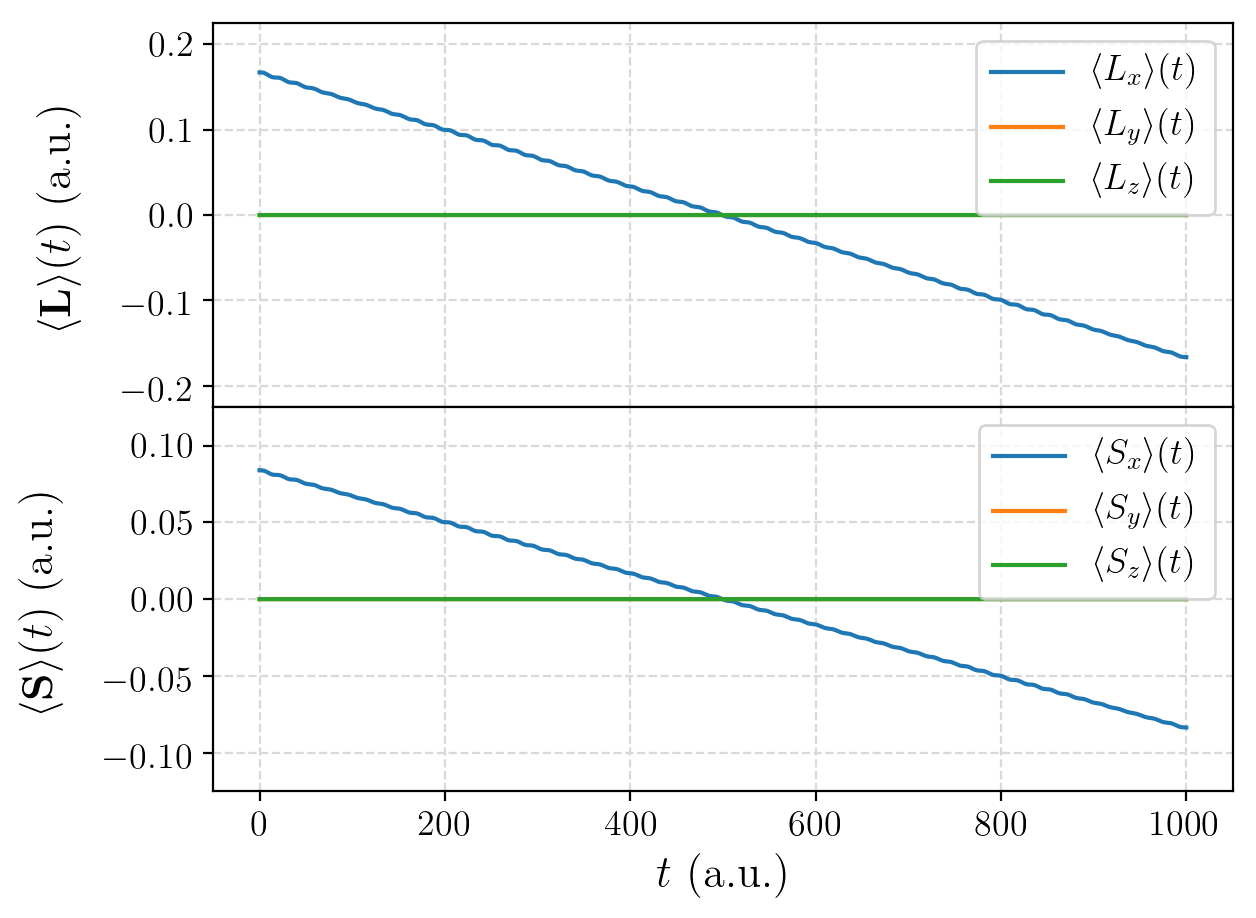}
		\caption{}
		\label{fig:strongSOC_B=-5000+10t_1}
	\end{subfigure}%
    \vskip\baselineskip
	\begin{subfigure}{.5\textwidth}
		\centering		\includegraphics[width=1.0\linewidth]{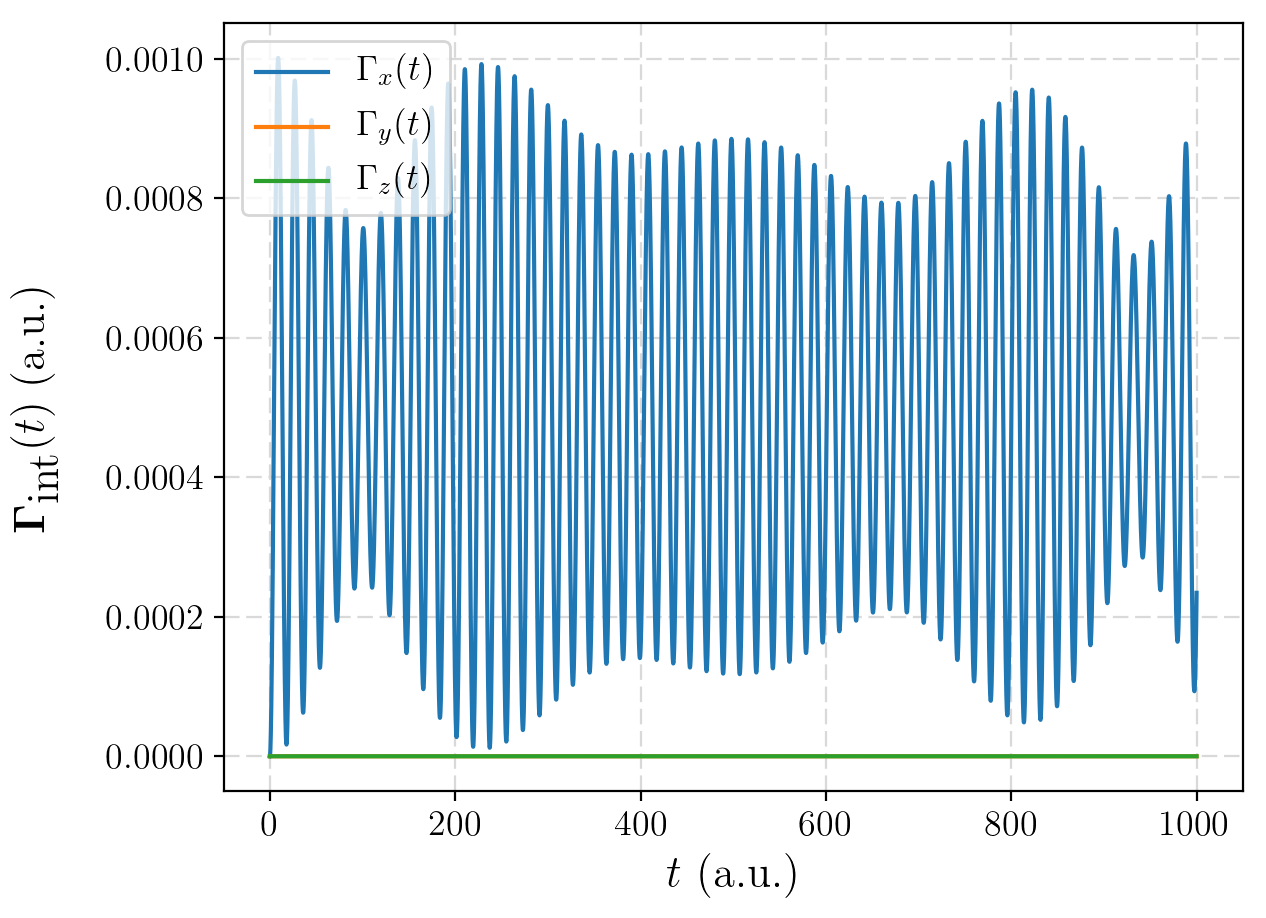}
		\caption{}
		\label{fig:strongSOC_B=-5000+10t_2}
	\end{subfigure}
    \caption{Time evolution of (a) the angular momenta and (b) the torque expectation values when $\xi=\SI{e3}{eV}$ and $B(t) = \num{-5e3}+10t$~T. The spin and orbital angular momenta are approximately proportional with a constant of proportionality of 2. Since SOC dominates the physics, the orbital and spin angular momenta are locked together and evolve in parallel. The orange lines are hidden beneath the green lines.}
	\label{fig:strongSOC_B=-5000+10t}
\end{figure}

The simulation results for the case of extremely strong SOC ($\xi = 10^3$ eV) are shown in Fig.~\ref{fig:strongSOC_B=-5000+10t}. The $(\xi/\hbar^2) \bm{L}\cdot\bm{S}$ spin-orbit term is now by far the largest in the Hamiltonian. Figure~\ref{fig:strongSOC_B=-5000+10t_1} shows that $\expval{\vec{L}}$ and $\expval{\vec{S}}$ are proportional to each other, with a proportionality constant of $1/2$.

This result can be understood by considering a Hamiltonian constructed solely from the SOC term. This decouples the two atoms since the SOC interaction acts on-site in our model. The Hamiltonian of one atom takes the form
    \begin{equation}
    	H = \frac{\xi}{\hbar^2} \vec{L}\cdot\vec{S} 
          = \frac{\xi}{2\hbar^2} (J^2 - L^2 - S^2),
    \end{equation}
where ${\xi=\num{e3}}$~eV. The orbital energies are $\frac{\xi}{2}(j(j+1)-l(l+1)-s(s+1)) = \frac{\xi}{2}(j(j+1)-2.75)$, with $j=1/2$ or $3/2$. The six eigenstates comprise a doublet with energy \SI{-e3}{eV}, corresponding to the $j=1/2$ states with $m_j = \pm 1/2$, and a quadruplet with energy \SI{+500}{eV}, corresponding to the $j=3/2$ states with $m_j=3/2$, $1/2$, $-1/2$, and $-3/2$. In the ground state of an oxygen atom (which contains 4 electrons in our $p$-band TB model), the two $j=1/2$ orbitals are filled first, leaving the other two electrons to occupy the four available $j=3/2$ orbitals. Using the Wigner-Eckart theorem~\cite{cohen-tannoudji}, it can be shown that the valence electrons residing in the $j=3/2$ states have $\bra{\phi}\vec{S}\ket{\phi}=\frac{1}{2}\bra{\phi}\vec{L}\ket{\phi}$ which explains the proportionality between $\expval{\vec{S}}$ and $\expval{\vec{L}}$ in Fig.~\ref{fig:strongSOC_B=-5000+10t_1}.


Above we noted the link between the torque exerted by the electrons on the nuclei, $\bm{\Gamma}_{\text{int}}$, and the rate of change of electronic angular momentum, $\expval{\vec{J}}$. Eq.~\eqref{eq:parallel_to_B_field} of appendix~\ref{app:torque} shows that their vector components parallel to the $B$ field are related by a minus sign. This link can be verified computationally by approximating $-{d\expval{\vec{J}}}/dt$ using finite differences of the total electronic angular momentum $\expval{\vec{J}}$ and comparing the result with the torque $\bm{\Gamma}_{\text{int}}$ acting on the nuclei. 

The torque on the nuclei due to $-{d\expval{\vec{J}}}/{dt}$ can be separated into orbital and spin contributions. The orbital contribution, $-{d \expval{\vec{L}}}/{dt}$, can be interpreted as in classical physics. The spin contribution, $-{d\expval{\vec{S}}}/{dt}$, corresponds to the EdH effect.
In the absence of SOC, as was shown in section~\ref{sec:nosoc}, the spin direction cannot reverse when the applied field reverses by changing its magnitude along a fixed axis. In this case $\unit{B}\cdot\bm{\Gamma}_\textrm{int} = -\unit{B}\cdot{d\expval{\vec{L}}}/{dt}$, implying that the EdH effect cannot occur without SOC.
For intermediate SOC strengths (Fig.~\ref{fig:intermediateSOC_B=-5000+10t}) we saw that the spin flip influences the direction of the torque. Both contributions to $\unit{B}\cdot\bm{\Gamma}_\textrm{int} = -\unit{B}\cdot{d \expval{\vec{L}}}/{dt} -\unit{B}\cdot{d \expval{\vec{S}}}/{dt}$ contribute appreciably in this case.
For extremely strong SOC strengths (Fig.~\ref{fig:strongSOC_B=-5000+10t}), $\expval{\vec{L}}$ and $\expval{\vec{S}}$ lock together and its average over a short time window does not change significantly as a function of time. 
The initial magnitude of $\expval{\vec{S}}$ is smaller than in the previous simulations, which is due to the $j=3/2$ eigenfunctions of the SOC term having $\bra{\phi}\vec{S}\ket{\phi} = \frac{1}{2}\bra{\phi}\vec{L}\ket{\phi}$ as is described above.

\section{Conclusions}\label{sec:conclusions}

This paper investigated a non-collinear TB model capable of simulating the EdH effect in a dimer. Since the EdH effect is based on the realignment of spins in a uniaxial $B$ field, it cannot occur in the absence of SOC. Based on simulation results showing the torque on the dimer for three very different SOC strengths, we showed how avoided crossings and the rate of change of magnetic field affect the reversal of electronic spins.
We also showed that the EdH effect is not the only source of torque on a dimer in a time-varying $B$ field: there is an additional contribution from the change in the electronic orbital angular momentum. The direct action on the nuclear charges of the Faraday electric field associated with the rate of change of the applied magnetic field also exerts a torque on the nuclei. This is very small for experimentally achievable values of $d\vec{B}/dt$ (although significant in our simulations).

Future work on this topic could aim to better understand the mechanism that translates the microscopic EdH effect to the better known macroscopic observations. How is the EdH effect modified by the quenching of $\vec{L}$ in systems without rotational symmetry, such as the iron clusters investigated in Ref.~\cite{Niemeyer2012}? What are the consequences of irreversibility and the loss of energy into other microscopic degrees of freedom? Now that the dimer model has been established, it can be scaled up to treat larger assemblies of atoms and a more diverse range of atom types. Another potentially fruitful area of investigation would be to look for effective classical models capable of emulating the effects of SOC and yet simple enough to be used in large-scale simulations of real materials problems, including studies of radiation damage in ferromagnetic steels.

\begin{acknowledgments}
This work was supported through a studentship in the Centre for Doctoral Training on Theory and Simulation of Materials at Imperial College London, funded by EPSRC grant EP/L015579/1. This work has been carried out within the framework of the EUROfusion Consortium and has received funding from the
Euratom research and training program 2019–-2020 under Grant Agreement No. 633053 and from the RCUK Energy Programme [Grant No. EP/P012450/1]. WMCF's visit to the University of Illinois at Urbana-Champaign, where much of his work on this project was done, was supported by the US Department of Energy under grant NA 0002911. We acknowledge support from the Thomas Young Centre under grant TYC-101.
\end{acknowledgments}
\appendix
\section{Relating the Dimer Torque to the Electronic Angular Momentum}\label{app:torque}

Below we show how the total torque on the dimer nuclei is a sum of the direct torque due to the interaction of the nuclei with the electromagnetic (EM) field and the torque due to the Coulomb attraction between the electrons and the nuclei. 

We consider a system of $n$ particles (nuclei and electrons), all of which have spins and interact with the externally applied EM four-potential, $(\varphi_a/c,\vec{A}_a) = (\varphi(\vec{r}_a,t)/c,\vec{A}(\vec{r}_a,t))$ where $c$ is the speed of light. The single-particle Hamiltonian of a particle in an EM field~\cite{white}, generalised to multiple interacting particles is,
\begin{align}
     H &= \sum_{a=1}^{n}\bigg[\frac{1}{2m_a}(\bm{p}_a - q_a\bm{A}_a)^2 + q_a\varphi_a \notag \\
     &\qquad\qquad- \frac{q_a\hbar}{2m_a} \bm{\sigma}_a\cdot\bm{B}_a + \frac{\xi_a}{\hbar^2} \vec{L}_a\cdot\vec{S}_a \bigg] \notag \\
     &\qquad\qquad+ \frac{1}{2}\sum_{a,b\neq a}^n V(\lvert \vec{r}_a - \vec{r}_b \rvert) ,
     \label{eq:Ham_EM}
\end{align}
where particle $a$ has mass $m_a$, charge $q_a$, SOC parameter $\xi_a$, and $V(r)$ is the interaction potential energy for a pair of particles separated by a distance $r$. 

By calculating the rate of change of the kinetic momentum $\bm{p}_{a} - q_{a} \bm{A}_{a}$, we obtain the Lorentz force operator for particle $a$:
\begin{align}
    \vec{F}^{EM}_a = \frac{q_a}{2} (\vec{v}_a\times\bm{B}_a -
    \bm{B}_a\times\vec{v}_a ) + q_a\bm{E}_a , \label{eq:nuclearEMforce}
\end{align}
where $\bm{E}_a = -\bm{\nabla}_a\varphi_a - \partial_t \vec{A}_a$ is the electric field operator on particle $a$ and $\vec{v}_a =  \frac{1}{m_a}(\vec{p}_a - q_a\vec{A}_a)$ is the velocity operator for particle $a$. Since we are working in the rest frame of the molecule, the nuclei have no orbital angular momentum and experience no spin-orbit coupling. Equation (\ref{eq:nuclearEMforce}) was derived quantum mechanically but reduces to the familiar classical Lorentz force law if the nuclei are treated classically, as is the case in this work. The direct electromagnetic torque on the classical nuclei is $\vec{\Gamma}^{N, EM} = \sum_{a \in N} \vec{r}_a \cross \vec{F}_a^{EM}$, where $\vec{r}_a$ is the classical position of nucleus $a$ and $\vec{F}_a^{EM}$ is the direct electromagnetic force on nucleus $a$. The total torque exerted on the set $N$ of nuclei in the molecule is given by
\begin{align}
   \vec{\Gamma}^{N} = \vec{\Gamma}^{N, EM} + \vec{\Gamma}_\textrm{int} ,
   \label{eq:true_mechanical_torque}
\end{align}
where $\vec{\Gamma}_\textrm{int} = \expval{\sum_{a \in N, b \in E} \vec{r}_a \cross \vec{F}_{ab}}$ is the Hellmann-Feynman torque exerted on the nuclei by the set $E$ of electrons, and $\vec{F}_{ab} = -\bm{\nabla}_a V(\lvert\vec{r}_a - \vec{r}_b\rvert)$ is the operator for the Hellman-Feynman force on nucleus $a$ due to electron $b$.

Let $\vec{J} = \sum_{a\in E} \vec{J}_a$ be the total electronic canonical angular momentum, where  $\vec{J}_a = \vec{L}_{a} + \vec{S}_{a}$ is the sum of the canonical orbital and spin angular momenta of electron $a$. The rate of change of the expectation value of $\vec{J}$ can be calculated using the Ehrenfest theorem:
\begin{align}
    \frac{d\expval{\vec{J}}}{dt} = \frac{1}{i\hbar}\expval{[\vec{J},H]}.
\end{align}
Simplifying the result by assuming that the applied $B$ field is uniform, working in the Coulomb gauge, and neglecting terms of second order in $\vec{A}$ yields
\begin{align}
    \frac{d}{dt} \expval{\vec{J}} = -\vec{\Gamma}_\textrm{int} - \vec{B}\cross\expval{\bm{\mu}},
\end{align}
and thus,
\begin{align}
   \unit{B}\cdot\frac{d}{dt} \expval{\vec{J}} &= -\unit{B}\cdot\vec{\Gamma}_\textrm{int},
   \label{eq:parallel_to_B_field}
\end{align}
where $\unit{B}$ denotes the unit vector pointing in the direction of the $B$ field.

\bibliography{\jobname} 

\end{document}